%% file: main.tex
 \documentclass[journal=cmatex,manuscript=article]{achemso}

\usepackage[version=3]{mhchem} 
\usepackage{subfig} %
\usepackage[dvipsnames]{xcolor}
\usepackage{booktabs}
\usepackage{multirow}

\author{Bruno Ipaves}
\affiliation{Instituto de F\'{\i}sica, Universidade de S\~ao Paulo, \\
CEP 05508-090, S\~ao Paulo - SP, Brazil}
\author{Jo\~ao F. Justo}
\affiliation{Escola Polit\'ecnica, Universidade de S\~ao Paulo, \\
CEP 05508-010, S\~ao Paulo - SP, Brazil}
\author{Lucy V. C. Assali}
\affiliation{Instituto de F\'{\i}sica, Universidade de S\~ao Paulo, \\
CEP 05508-090, S\~ao Paulo - SP, Brazil}
\email{ipaves@if.usp.br, jjusto@lme.usp.br, lassali@if.usp.br}

\title[An \textsf{achemso} demo]
{Functionalized few-layer silicene nanosheets: density functional theory on stability, structural, and electronic properties}

\begin{document}

\input{tocentry/tocentry.tex}
\input{abstract/abstract.tex}
\input{introduction/introduction.tex}

\input{methodology/methodology.tex}

\input{results/results.tex}
\input{discussion/discussion.tex}

\input{acknowledgement/acknowledgement.tex}

\bibliography{references}

\end{document}

%% file: tocentry/tocentry.tex
\begin{tocentry}





\includegraphics[scale = 0.1, trim={0cm 0.0cm 0cm 0cm}, clip]{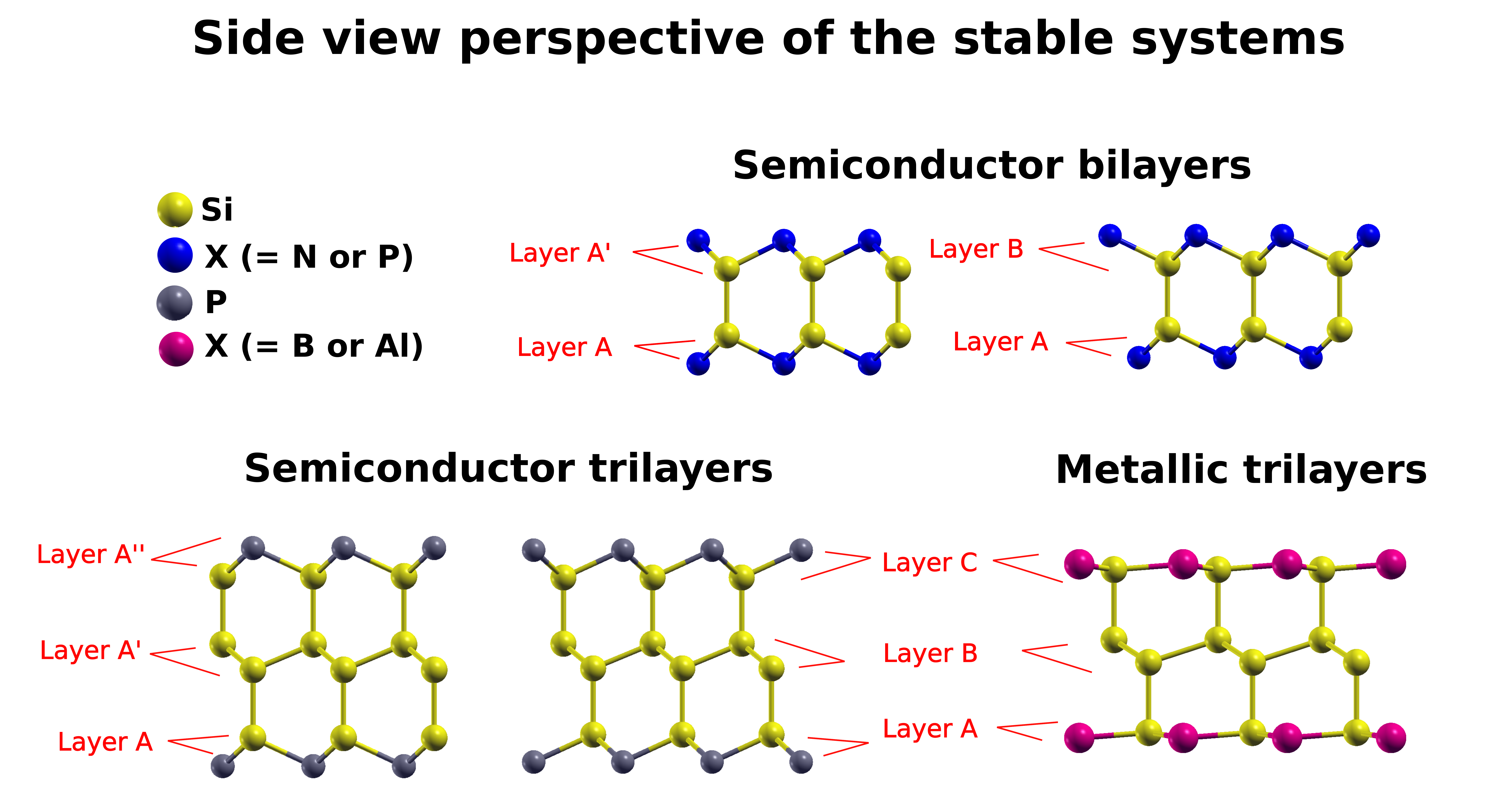}

\end{tocentry}

%% file: abstract/abstract.tex
\begin{abstract}
Using density functional theory calculations, we investigated the properties of few-layer silicene nanosheets, namely bilayers and trilayers, functionalized with group-III or group-V atoms of the periodic table. We considered the \ce{Si2X2} bilayers and the \ce{Si2X4} trilayers, X = B, N, Al, P. We computed the structural, energetic, dynamic, elastic, and electronic properties of those systems in several stacking configurations, labeled as \ce{AA$'$}, \ce{AB}, \ce{AA$'$A$''$}, and \ce{ABC}. The results revealed that \ce{AA$'$-Si2N2}, \ce{AB-Si2N2}, \ce{AA$'$-Si2P2}, \ce{AB-Si2P2}, \ce{ABC-Si4B2}, \ce{ABC-Si4Al2}, \ce{AA$'$A$''$-Si4P2}, and \ce{ABC-Si4P2} nanosheets are all dynamically stable, according to their respective phonon dispersion spectra. Additionally, by comparing the standard enthalpies of formation of doped few-layer silicene systems with the ones of the pristine silicene monolayer, bilayer, and trilayer nanosheets, we found that those structures could be experimentally accessed. Exploring the electronic properties of those stable systems, we discovered that the silicene bilayers and trilayers functionalized with N or P atoms change from a metallic to a semiconducting behavior. However, the metallic behavior is kept when the nanosheets are functionalized with B or Al atoms. Finally, by placing our results within the context of silicon-based systems previous investigations, we could envision potential applications for those nanosheets in van der Waals heterostructures, alkali-metal ion batteries, UV-light devices, and thermoelectric materials.
\end{abstract}
\eject

%% file: introduction/introduction.tex
\section{Introduction}

Since the discovery and isolation of graphene sheets, \cite{novoselov2004} several 2D materials have been reported in the literature with different physical properties, with metallic (\ce{NbSe2}), insulating (\ce{hBN}), and semiconducting (\ce{MoS2}) behaviors \cite{novoselov20162d,li2018epitaxial}. Due to their remarkable properties, those 2D materials have been considered for applications in many areas, such as in electronics, medicine, energy generation and storage, light processing, and sensors and actuators. 

Silicene, a 2D hexagonal silicon monolayer, presents some properties analogous to those of graph\-ene, including the honeycomb structure with Dirac cone at the high-symmetry K-point in the first Brillouin zone \cite{cahangirov2009two,garcia2011,schoop2018chemical}. Hence, it has been considered a promising candidate to be used in similar applications like those of graphene \cite{molle2018silicene}. Unlike graphene, silicene is not a flat nanosheet due to the silicon $sp^2$-$sp^3$ hybridizations, resulting in a low-buckled structure \cite{garcia2011,zhuang2015honeycomb,schoop2018chemical}. This buckling makes silicene very reactive to hydrogenation, fluorination, and doping with substitutional atoms, which modify substantially its properties and may transform it from a zero band gap semimetal into a metallic or semiconducting material \cite{sivek2013adsorption, ding2013density}. Although silicene monolayer has been intensively studied, systems involving few-layer silicene (FLS) systems have received less attention over the last few years \cite{qian2020multilayer}. However, it is considerably easier to synthesize FLS than the silicene monolayer \cite{liu2013structural}. Furthermore, FLS properties are highly dependent on the stacking order and interlayer interactions, e.g., the silicene bilayer could have metallic or semiconducting behavior by simply changing its morphology \cite{padilha2015free,fu2014stacking}, a feature that is very attractive from a technological perspective.

Free-standing FLS nanosheets have been synthesized by liquid oxidation and exfoliation of \ce{CaSi2} and three possible stacking configurations have been proposed: AA, AABB, and ABC \cite{liu2018few}. There are recent studies of silicon-based binary compounds, such as \ce{Si2X2} (X = N, P, As, Sb, or Bi) \cite{ozdamar2018structural} and \ce{Si2XY} (X, Y = P, As, Sb, or Bi), which exhibit a variety of properties \cite{touski2021structural}. These 2D binary systems are structurally similar to the silicene bilayer since they are composed of two silicene sheets 50\% doped with substitutional atoms, as Si-Si bonds between neighboring layers are covalent. The 2D silicon-based materials have potential applications in several fields, such as anodes in lithium-ion batteries (LIBs) \cite{liu2018few}, spintronics \cite{touski2021structural}, optoelectronic and electronic  devices \cite{yang2016two}, energy storage \cite{an2020recent}, and field-effect transistor pressure sensors \cite{tantardini2021computational}.

Considering the potential applications of FLS nanosheets, it is of great interest to understand the properties of these materials in several functionalized structures. Here, we performed a theoretical investigation on the physical properties of silicene bilayers and trilayers functionalized with group-III (B and Al) and group-V (N and P) atoms of the periodic table in a variety of stacking configurations, using the density functional theory and the supercell approach. We computed the structural and electronic properties of pure FLS and the changes resulting from such functionalizations. Then, we determined the dynamic stability, enthalpy of formation, and elastic constants of those structures. 

We labeled the functionalized silicene bilayers and trilayers respectively as \ce{Si2X2} and \ce{Si4X2} (X = B, N, Al, P). The bilayers consist of two 50\% X-doped silicene nanosheets and the trilayers consist of three silicene nanosheets, one non-doped silicene layer between two 50\% X-doped ones. Herein, we considered four stacking configurations: \ce{AA$'$} and \ce{AB} for bilayers; and \ce{AA$'$A$''$} and \ce{ABC} for trilayers, with Si-Si covalent bonds. We found that the \ce{Si2B2} and \ce{Si2Al2} bilayers, and the \ce{AA$'$A$''$-Si4N2}, \ce{ABC-Si4N2}, \ce{AA$'$A$''$-Si4B2}, and \ce{AA$'$A$''$-Si4Al2} trilayers are not dynamically stable. We explored in depth the physical properties of dynamically stable nanosheets, that can be classified in two distinct groups according to their band structure: (i) the  metallic \ce{ABC-Si4B2} and \ce{ABC-Si4Al2} trilayers; (ii) the indirect band gap semiconductors  \ce{Si2N2} and \ce{Si2P2} bilayers in both \ce{AA$'$} and \ce{AB} stacking configurations, as well as the \ce{Si4P2} trilayer in both \ce{AA$'$A$''$} and \ce{ABC} stacking arrangements.

Regarding the dynamically stable systems, they present enthalpies of formation lower than that of pristine silicene and, hence, they could be synthesized easily. Particularly, we found that the indirect band gap semiconductors are the most promising systems to be produced as free-standing 2D materials. Finally, based on our results combined with the ones from previous works that studied similar structures, we suggest that these stable systems could be explored for van der Waals heterostructures, alkali-metal ion batteries (AMIBs), UV-light devices, and thermoelectric materials applications.   

The paper is organized as follows: first, we present the theoretical methodology and computational details, then the results on energetic, structural, stability, and electronic properties are presented. The last section presents a discussion and concluding remarks.

%% file: methodology/methodology.tex
{\section{Theoretical Model and Computational  Details}}

All calculations were performed using the Quantum ESPRESSO computational package \cite{Giannozzi2009,giannozzi2017advanced} with the electronic interactions described within the density functional theory \cite{hohenberg1964inhomogeneous, kohn1965self} (DFT) and the wave functions were expanded with the projector augmented wave (PAW) method \cite{Kresse}. We used the generalized gradient approximation of Perdew-Burk-Ernzerhof (PBE) for the exchange and correlation potential \cite{perdew1996generalized}, augmented by the dispersive van der Waals interaction within the Dion {\it et al.} scheme \cite{dion2004} and optimized by Klime{\v{s}} {\it et al.} (optB88-vdW) \cite{klimevs2009}. Those approximations have been used by recent investigations, which have shown the relevance of describing the vdW interactions appropriately, particularly when treating the vibrational properties of the systems \cite{zhang2011van, park2015van, marcondes2018importance}. 

An 1100 eV energy cutoff for the valence electrons plane wave expansion was used and self-consistent iterations were performed until reaching convergence in total energies of 0.1 meV/atom. During structural optimization, relaxations and distortions were considered in all ions, without symmetry constraints, until forces in any ion were lower than 1 meV/{\AA}. The irreducible Brillouin zones (BZ) for computing the electronic states were sampled by a $16 \times 16 \times 1$ Monkhorst-Pack $k$-point grid \cite{Monkhorst}.

The 2D structures were built with a hexagonal simulation supercell, considering periodic boundary conditions. A lattice parameter of 25 {\AA} was used in the perpendicular direction to the sheets ($z$-axis) to prevent interactions among cell images. The primitive cells contained 4 and 6 atoms, respectively,  for the bilayers and  trilayers. The cell parameters in the $xy$ plane were obtained using a variable-cell optimization, and the respective phonon dispersion curves were obtained through the density functional perturbation theory (DFPT) \cite{baroni2001phonons}, with the irreducible Brillouin zones sampled by an $8 \times 8 \times 1$ $q$-point mesh.

To determine the elastic properties of the systems, we built a rectangular cell with 8 atoms for the bilayers and 12 atoms for the trilayers. The elastic constants were evaluated with the strain–energy method, by applying two in-plane strains ($\epsilon$), ranging from -1.5\% to 1.5\% with respect to the optimized cell parameters: (i) uniaxial deformation along the zigzag direction and (ii) biaxial planar deformation (zigzag and armchair simultaneously) \cite{cadelano2010elastic}. The elastic energy is expressed by  
\begin{equation}
  E(\epsilon) = E(0) + \frac{1}{2}U^{(2)} \epsilon^{2} + O(\epsilon^3),
  \label{eq_elastic_energy}
\end{equation}
where $E(\epsilon)$ and $E(0)$ are the total energies of strained and unstrained configurations, respectively. For isotropic structures, $U^{(2)}$ is the $\rm C_{11}$ elastic constant for zigzag axial deformation, whereas it is $\rm 2(C_{11} + C_{12})$ for biaxial planar deformation \cite{cadelano2010elastic}. Accordingly, the relevant elastic constants were obtained by fitting a second-order polynomial to the data.

The standard enthalpy of formation $\Delta H_{\!f}^0$, per atom and at zero temperature, of each structure was computed by
\begin{equation}
 \Delta H_{\!f}^0 = \frac{E_t({\rm Si}_y{\rm X}_2) - yE_t({\rm Si}) - 2E_t({\rm X})}{y + 2},
 \label{eq_formation_energy}
 \end{equation}
where $E_t({\rm Si}_y{\rm X}_2)$ is the total energy of the 2D system, with $y$ Si atoms and 2 X atoms (X = B, N, Al, P), $y$ being 2 and 4 for bilayers and trilayers, respectively. Since the standard enthalpy of formation is a measurement of the energy released or absorbed when a substance is synthesized from its pure elements, the $E_t$(Si) and $E_t$(X) are the total energies, per atom, of the respective Si and X standard states. Those energies, computed within the same methodology previously described, were obtained from the total energy of crystalline silicon in a diamond lattice, boron in a trigonal crystalline structure ($\beta$-boron), aluminum in a face-centered cubic crystalline structure, nitrogen in an isolated N$_2$ molecule, and bulk black phosphorus. This procedure to determine enthalpies and/or energies of formation has been successfully used to investigate several other systems \cite{assali2006manganese,assali20113,haastrup2018computational}.

%% file: results/results.tex
\section{Physical Properties of Functionalized FLS Nanosheets}

\subsection{Structural properties}

\begin{figure}[hp]
\centering
\includegraphics[scale = 0.13, trim={0cm 0cm 0cm 0cm}, clip]{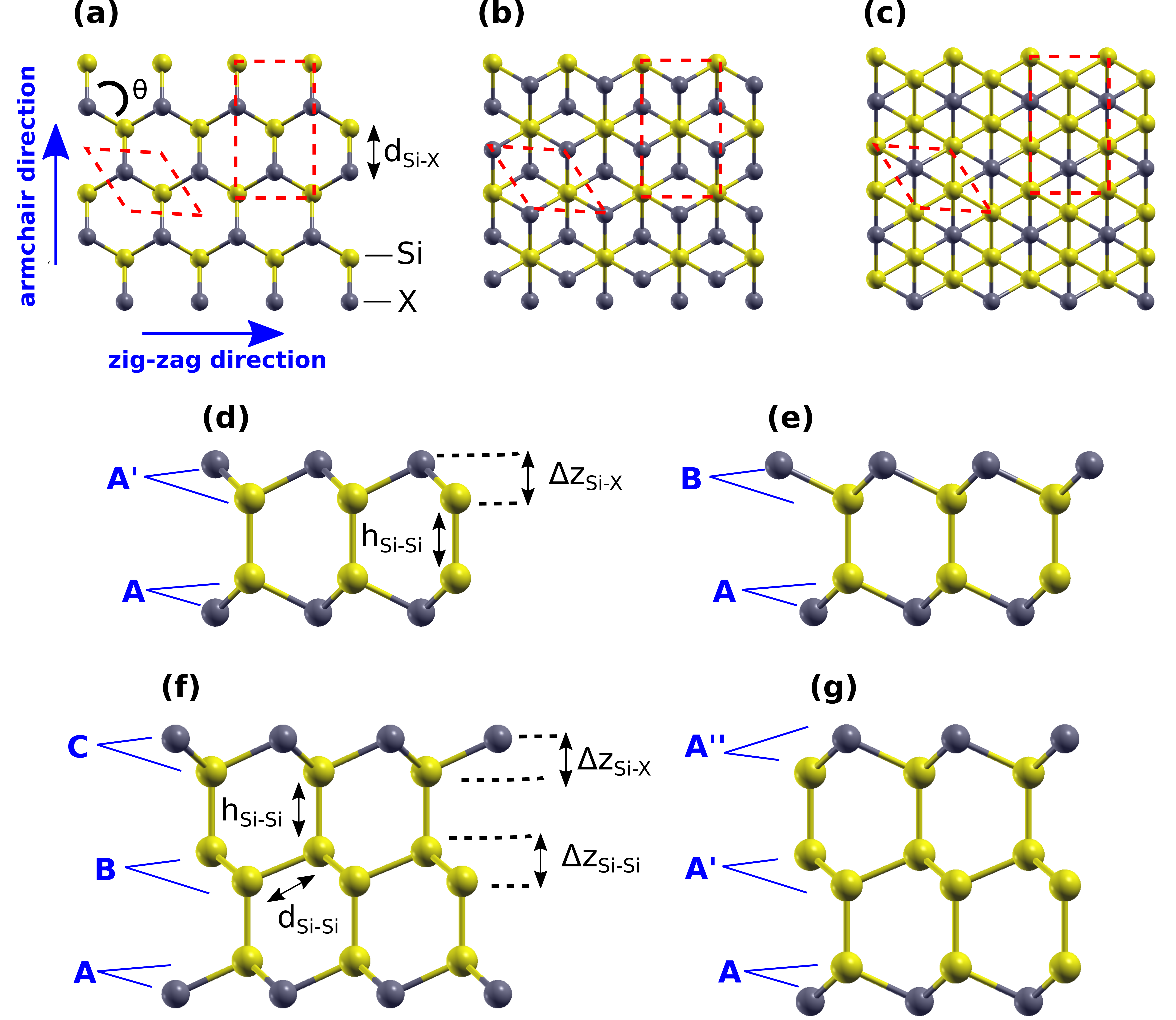}
\vspace{0.25cm}
\caption{Schematic representations of the \ce{Si2X2} bilayers and \ce{Si4X2} trilayers (X = B, N, Al, P) structures. Herein, X = Si means pristine silicene bilayer (Si$_4$) and trilayer (Si$_6$). Top view perspective of (a) \ce{AA$'$} and \ce{AA$'$A$''$}, (b) \ce{AB}, and (c) \ce{ABC} stacking configurations; and side view perspective of (d) \ce{AA$'$}, (e) \ce{AB}, (f) \ce{ABC}, and (g) \ce{AA$'$A$''$} stacking configurations. The \ce{AA$'$A$''$} stacking is also known as \ce{AA$'$A} in pure silicene trilayer. The yellow and gray spheres represent respectively Si and X atoms. The red dashed lines represent the unit cell limits, hexagonal (rectangle) cells used to determine dynamic, structural, and electronic (elastic) properties. These schemes also indicate the labels given to the $d_{\rm Si-Si}$ and $d_{\rm Si-X}$ intralayer, $h_{\rm Si-Si}$ interlayer, $\Delta z_{\rm Si-Si}$ and $\Delta z_{\rm Si-X}$ buckling distances, and the $\theta$ intraplanar bond angle.}
\label{schematic_representation}
\end{figure}

Figure \ref{schematic_representation} shows schematic representations of the \ce{Si_2X_2} bilayers, in the \ce{AA$'$} and \ce{AB} stacking arrangements, and \ce{Si_4X_2} trilayers, in  the \ce{AA$'$A$''$} (or \ce{AA$'$A}) and \ce{ABC} stacking configurations, with X = B, N, Al, P. Moreover, when X = Si in the figure, it indicates pristine  silicene monolayer \ce{Si2},  bilayer \ce{Si4}, and trilayer \ce{Si6}, which received those labels due to the number of Si atoms in the simulation unit cell. The figure also exhibits the labels given to the intralayer ($d$), interlayer ($h$), and buckling ($\Delta z$) distances, as well as the intraplanar bond angle ($\theta$).

Table \ref{table_properties} presents the optimized structural parameters of the \ce{Si2X2} bilayers and \ce{Si4X2} trilayers, as well as the respective values obtained for pristine silicene Si$_2$ monolayer, Si$_4$ bilayers, and Si$_6$ trilayers, where the distance labels are consistent with the ones defined in figure \ref{schematic_representation}. The table also displays the respective standard enthalpy of formation $\Delta H_{\!f}^{0}$ and the dynamic stability (DS).

\begin{table}[hp]
\begin{center}
\begin{footnotesize}
\caption{Structural properties of \ce{Si2X2} bilayers and \ce{Si4X2} trilayers (X = B, N, Al, P): lattice parameters ($a$), intralayer ($d$), interlayer ($h$), and buckling ($\Delta z $) distances, given in {\AA}, and the intraplanar bond angle ($\theta$), labeled according to figure \ref{schematic_representation}. The standard enthalpy of formation ($\Delta H_{\!f}^{0}$), obtained according to equation \ref{eq_formation_energy}, is given in meV/atom. For pristine silicene Si$_2$ monolayer, Si$_4$ bilayers, and Si$_6$ trilayers,  X = Si. The last column indicates the dynamic stability (DS) of the structures.} 
\vspace{0.4cm}
\begin{tabular}{lllcccccccrl}
\cmidrule{4-10}\morecmidrules\cmidrule{4-10}
&  &  &  &  & &  &  & & &\\[-5mm]
\multicolumn{3}{c}{}   &   \multicolumn{6}{c}{Structural parameters} & \multicolumn{1}{c}{} \\
&  &  &  &  & &  &  & & &\\[-4mm]
\hline \hline
&  &  &  &  & &  &  & & &\\[-4mm]
\multicolumn{1}{l}{System}  & \multicolumn{1}{c}{Structure}  & 
\multicolumn{1}{l}{Stacking} & $a$  & $d_{\mathrm{Si-X}}$  & $h_{\mathrm{Si-Si}}$ &  $d_{\mathrm{Si-Si}}$ & $\Delta z_{\mathrm{Si-X}}$  & $\Delta z_{\mathrm{Si-Si}}$ & $\theta $ & $\Delta {H}_{\!f}^0$ & DS\\ 
 &  &  &  &  & &  &  & & &\\[-4mm]
\hline \hline
 &  &  &  &  & &  &  & & &\\[-4mm]
Monolayer~~ & \multicolumn{1}{l}{Si$_2$} & --- & 3.870 & 2.285 &  --- & --- & 0.482 & --- & 115.67 & 751 & Yes \\  
\hline
 &  &  &  &  & &  &  & & &\\[-4mm]
\multirow{10}{*}{Bilayer} & \multirow{2}{*}{Si$_4$} & AA$'$ & 3.854 & 2.325 &  2.470 & ---  &  0.676 &  --- & 111.89 & 609 & Yes \\ 
   &     &  AB   &  3.847 & 2.322 & 2.538   & ---&  0.676 & ---  & 111.89 & 570 & Yes  \\  \cline{2-12} 
 &  &  &  &  & &  &  & & & & \\[-4mm]
   & \multirow{2}{*}{Si$_2$B$_2$}  & AA$'$ & 3.448 & 1.993 & 2.565 & --- & 0.109 & ---& 119.70 & 648 &  No \\ 
    &     &  AB & 3.468 & 2.005 &  2.498 & --- & 0.104 & --- & 119.73 & 704 & No\\  \cline{2-12} 
 &  &  &  &  & &  &  & & &\\[-4mm]
  &\multirow{2}{*}{Si$_2$Al$_2$}& AA$'$  & 4.231 & 2.457 & 2.572 & --- & 0.271 & ---& 118.79 & 446 & No \\
  &     &   AB  & 4.242 & 2.457 &  2.514 & --- & 0.200 & --- & 119.34 & 404 & No\\   \cline{2-12}
 &  &  &  &  & &  &  & & & \\[-4mm]
&\multirow{2}{*}{Si$_2$N$_2$}&  AA$'$  & 2.909 & 1.767 & 2.429 & --- & 0.548 & ---& 110.81 & -743 & Yes \\ 
 & & AB & 2.921 & 1.771 &  2.393 & --- & 0.541 & --- & 111.08 & -733 & Yes\\   \cline{2-12}
 &  &  &  &  & &  &  & & &\\[-4mm]
 &\multirow{2}{*}{Si$_2$P$_2$}& AA$'$ & 3.544 & 2.286 & 2.380 & --- & 1.019 & ---& 101.65 & -87 & Yes \\
  & & AB & 3.557 & 2.287 &  2.351 & --- & 1.007 & ---& 102.06 &  -84 & Yes\\   \hline
 &  &  &  &  & &  &  & &\\[-4mm]
\multirow{10}{*}{Trilayer} & \multirow{2}{*}{Si$_6$} & AA$^{\prime}$A & 3.858 & 2.332 & 2.423 & 2.355 & 0.690  & 0.756   & 111.62 & 224 & Yes\\
& & ABC & 3.858 & 2.333 & 2.435   & 2.355 & 0.695  & 0.764   & 111.51 & 220 & Yes \\ \cline{2-12}
 &  &  &  &  & &  &  & & &\\[-4mm]
& \multirow{2}{*}{Si$_4$B$_2$} & AA$^{\prime}$A$^{{\prime}{\prime}}$  & 3.539 & 2.045 & 2.495 & 2.282 & 0.092  & 1.019  & 119.79 &  552 & No \\
 & & ABC & 3.568 & 2.061 & 2.403  & 2.272 & 0.086  & 0.900  & 119.82 & 529 & Yes\\ \cline{2-12}

 &  &  &  &  & &  &  & & &\\[-4mm]
& \multirow{2}{*}{Si$_4$Al$_2$}& AA$^{\prime}$A$^{{\prime}{\prime}}$  & 4.128 & 2.390 & 2.439 & 2.515 & 0.175  & 0.802 & 119.46 &  339 & No\\
 &  &  ABC & 4.143 & 2.399 & 2.419  & 2.516 & 0.185  & 0.782  & 119.40 & 309 & Yes\\ \cline{2-12}
&  &  &  &  & &  &  & & &\\[-4mm]
& \multirow{2}{*}{Si$_4$N$_2$} & AA$^{\prime}$A$^{{\prime}{\prime}}$ & 2.911 & 1.763 & 2.384 & 2.508& 0.533  & 1.861   & 111.27 & -156 & No\\
& & ABC & 2.865 & 1.751 & 2.433  & 2.807 & 0.574  &2.267 & 109.78 & -188 & No \\ \cline{2-12}
&  &  &  &  & &  &  & & &\\[-4mm]
& \multirow{2}{*}{Si$_4$P$_2$} & AA$^{\prime}$A$^{{\prime}{\prime}}$ & 3.641 & 2.321 & 2.365 & 2.283& 0.984  & 0.890   & 103.32 & -6 & Yes\\
 & & ABC  & 3.665 & 2.330 & 2.356  & 2.292 & 0.974  & 0.880   & 103.74 & -7 & Yes\\
\hline \hline
\label{table_properties}
\end{tabular}
\end{footnotesize}
\end{center}
\end{table}

Starting with the pristine FLS, we found that the lattice constants for all of nanosystems are very close, ranging from 3.847 {\AA} (\ce{AB-Si4}) to 3.870 {\AA} (\ce{Si2}). On the other hand, the calculated buckling distances in the \ce{Si4} and \ce{Si6} layers, despite having similar values, are on the order of 43\% larger than the value of 0.482 {\AA} found in pure silicene \ce{Si2} monolayer. Additionally, \ce{Si2} has an intraplanar bond angle $\theta$ of 115.67$^{\circ}$, which is related to $sp^2$-$sp^3$ hybridization, while the $\theta$ values, obtained for \ce{AA$'$A-Si6} and \ce{ABC-Si6} trilayers, are 111.62$^{\circ}$ and 111.51$^{\circ}$, respectively, which are closer to the $sp^3$ hybridization angle value of 109.47$^{\circ}$. As the number of layers increases, the $\theta$ values decrease towards the tetrahedral value, as Si atoms prefer the $sp^3$ hybridization \cite{li2019hydrogenation,hess2021bonding}. Moreover, unlike the few-layer graphene and graphite, which present weak interlayer van der Waals interactions, in the FLS the layers are covalently bonded. Accordingly, the  interlayer $h_{\rm Si-Si}$  distances are in the 2.423-2.538 {\AA} range, the smallest one associated with the trilayers, all values close to the interatomic distance of crystalline silicon (2.352 {\AA}). Those results are in good agreement with the ones reported in the literature \cite{kamal2013silicene, padilha2015free, li2019hydrogenation, qian2020multilayer}.

According to table \ref{table_properties}, when the silicene bilayer is 50\% doped with substitutional B, Al, or P atoms, the lattice parameter $a$ changes by about 10\%, decreasing in the \ce{Si2B2}  and \ce{Si2P2}  bilayers, and increasing in the \ce{Si2Al2} ones, regardless of the stacking configuration. In contrast, the N atom substitution considerably affects the $a$ value, i.e., the lattice parameter decreases $\approx$ 25\% in both \ce{Si2N2} stacking configurations. Therefore, while the Al substitution stretched out the lattice parameter $a$ and the intralayer distance $d_{\rm Si-Al}$, in comparison with pristine silicene bilayers, the incorporation of B, N, or P atoms reduced them. The values $d_{\rm Si-Al}=2.457$ {\AA} and $d_{\rm Si-B} \approx 2.0$ {\AA} for, respectively, \ce{Si2Al2}  and \ce{Si2B2} bilayers are close to the ones reported by Ding {\it et al.} \cite{ding2013density} in the Al/B doped silicene monolayer: 2.42 {\AA} for \ce{SiAl}, 2.43 {\AA} for \ce{AlSi3}, 1.96 {\AA} for \ce{SiB}, and 1.94 {\AA} for \ce{BSi3}. 

It can be noticed that in the bilayer systems functionalized with group-III atoms (B and Al, with valence electronic configuration $ns^2np^1,\, n=2$ and $3$, respectively), the intraplanar bond angles are greater than in the pristine \ce{Si4} bilayer, being very close to the value of 120$^{\circ}$, which is the expected value of the $sp^2$-type hybridization of B and Al atoms bonded to three adjacent Si atoms. Consequently, the  buckling distances are greatly affected by the substitution, being reduced by about 85\% in the \ce{Si2B2} bilayer for both stacking configurations ($\Delta z_{\rm Si-B} \approx 0.11$ {\AA}),  by about 60\% in the \ce{AA$'$-Si2Al2} bilayer ($\Delta z_{\rm Si-Al} \approx 0.27$ {\AA}),  and by about 70\% in the \ce{AB-Si2Al2} bilayer ($\Delta z_{\rm Si-Al} \approx 0.20$ {\AA}). 
 
 On the other hand, in the bilayer systems functionalized with group-V atoms (N and P, with valence electronic configuration $ns^2np^3,\, n=2$ and $3$, respectively), according to table \ref{table_properties}, the intraplanar bond angles are smaller than in the pristine \ce{Si4} sheet, approaching the $sp^3$-type hybridization angle value, with three bonds and a lone pair. Those values are a little greater than 109.47$^{\circ}$ for the \ce{Si2N2} systems while they are smaller than that value for the \ce{Si2P2} nanosheets. The P atom substitution causes major modifications in the pristine buckling distances as $\Delta z_{\rm Si-P} \approx 1.0$ {\AA} in the \ce{Si2P2} bilayer, for both stacking configurations, whilst for the N atom substitution the buckling distances are $\Delta z_{\rm Si-N} \approx 0.5$ {\AA}, a value that is in the range between the ones found for pristine silicene monolayer and bilayers. The functionalized bilayers present interlayer distances $h_{\rm Si-Si}$ slightly greater/smaller than the ones of pristine silicene bilayers, indicating that the strong Si-Si covalent bonds are maintained. Moreover, the structural parameters of \ce{Si2N2} and \ce{Si2P2} systems agree with the available reported data \cite{ozdamar2018structural}.
 
The pristine silicene \ce{Si6} trilayers have lattice parameters $a$ of 3.858 {\AA} and  present additional structural parameters in comparison to the bilayer systems, namely, the intralayer $d_{\rm Si-Si}$ and buckling $\Delta z_{\rm Si-Si}$ distances. The $d_{\rm Si-Si}=2.355$ {\AA} obtained for both stacking configurations, as well as the $\Delta{z_{\rm Si-Si}}$ of 0.756 {\AA} and 0.764 {\AA}, found in the \ce{AA$'$A} and \ce{ABC} stacking configurations, are very close to the Si bulk values of 2.352 {\AA} and 0.784 {\AA} for $d_{\rm Si-Si}$ and $\Delta{z_{\rm Si-Si}}$, respectively. 

The B, Al, or P doping of \ce{Si6} modify the lattice parameter by about 7\%, which increases in the \ce{Si4Al2} sheets and decreases in the \ce{Si4B2} and \ce{Si4P2} ones, irrespective of the stacking configuration. The \ce{Si4P2} structures exhibit the closest lattice parameters to the one of pure silicene trilayers, in which we found lattice parameters $a$ of 3.641 and 3.665 {\AA} for \ce{AA$'$A$''$} and \ce{ABC} stacking configurations, respectively. Nevertheless, the presence of N affects substantially the $a$ values, decreasing them by more than 25\%, as compared to that of the pristine systems, in both \ce{Si4N2} stacking configurations. Those systems present the smallest lattice parameters among all trilayers studied here, consistent with the \ce{Si2N2} systems among the bilayers.

The \ce{Si4Al2} trilayers present lattice parameters $a$ of about 4.1 {\AA}, with buckling distances $\Delta z_{\rm Si-Al}$ and intraplanar bond angles $\theta$ of about 0.18 {\AA} and 119.4$^{\circ}$, respectively. The \ce{Si4B2} trilayers present $a \approx 3.5$ {\AA}, $\Delta z_{\rm Si-B} \approx 0.09$ {\AA}, and $\theta \approx 119.8^{\circ}$. As a result, those nanosheets doped with group-III atoms present low-buckled surfaces, as expected for $sp^2$-type hybridization in a Si host. In the trilayers functionalized with N or P atoms, the intraplanar bond angles $\theta$ and the intralayer $d_{\rm Si-X}$ and buckling $\Delta z_{\rm Si-X}$ distances show similar behavior to the functionalized bilayers doped with group-V atoms. Accordingly, those nanosheets present buckled surfaces.

Regarding \ce{AA$'$A$''$-Si4Al2} and \ce{ABC-Si4Al2} systems, the intralayer distances are, respectively,  $d_{\rm Si-Al}$ = 2.390 and 2.399 {\AA}, which are slightly smaller than the values found for the \ce{Si2Al2} bilayers. Although the Si-Si intralayer distance $d_{\rm Si-Si}$ of around 2.52 {\AA} is larger than the one in pure silicene trilayer (2.355 {\AA}), silicon bulk (2.352 {\AA}), or silicene monolayer (2.285 {\AA}), the bond distance is still smaller than the longest Si-Si $\sigma$ bonds (2.697 and 2.7288(15) {\AA}) \cite{kyushin2020silicon}. These systems also present a small variation of the buckling  $\Delta z_{\rm Si-Si}$ distances, with values of 0.802 {\AA} for \ce{AA$'$A$''$-Si4Al2} and 0.782 {\AA}  for \ce{ABC-Si4Al2}. Additionally, the  $h_{\rm Si-Si}$ distance between layers, i.e., the Si-Si interlayer distance, is similar to that in pristine silicene trilayes (2.423 and 2.435 {\AA}) and close to the value of 2.365 {\AA} obtained for the hydrogenated silicene \cite{liu2013structural}, showing that the Al incorporation does not disrupt the strong covalent Si-Si interactions. 

As in the \ce{Si4Al2} systems, the functionalized \ce{Si4B2}, \ce{Si4N2}, and \ce{Si4P2}  trilayers have strong Si-Si covalent bonds, presenting $h_{\rm Si-Si}$ interlayer distance values close to the one in pristine  silicene trilayers. Moreover, except for the \ce{Si4N2} system, these structures present Si-Si intralayer distances close to the value in pure silicene monolayer of 2.285 {\AA}, being 2.272 and 2.282 {\AA} for \ce{ABC$-$} and \ce{AA$'$A$''$}\ce{-Si4B2} systems, respectively, while they are 2.292 and 2.283 {\AA} for \ce{ABC$-$} and \ce{AA$'$A$''$}\ce{-Si4P2} systems. In contrary,  the $d_{\rm Si-Si}= 2.807$ {\AA} found for \ce{ABC-Si4N2} trilayer is larger than the longest known Si-Si $\sigma$ bonds \cite{kyushin2020silicon}.

The buckling Si-Si distance $\Delta z_{\rm Si-Si}$ of 0.756 {\AA} and 0.764  {\AA} for pristine silicene trilayers in the \ce{AA$'$A} and \ce{ABC} stacking configurations, respectively, are very close to the Si bulk value of 0.784 {\AA}. For the trilayer doped with Al, this buckling distance does not increase more than 6\%, depending on the stacking configuration, when compared with  pristine silicene. For the \ce{Si4P2} trilayers, regardless the stacking configuration, the $\Delta z_{\rm Si-Si}$ distance increase is of about 17\%, of the same magnitude observed for the \ce{Si4B2} system in the \ce{ABC} stacking, which, however, is strongly affected, increasing by more than 34\% when the stacking is \ce{AA$'$A$''$}.  On the other hand, the N atom substitution causes huge modifications in the pristine Si-Si buckling distances, i.e., $\Delta z_{\rm Si-Si}$ increases about 150\% for \ce{AA$'$A$''$-Si4N2} stacking arrangement and has its value practically tripled for \ce{ABC} stacking configuration. Those major changes have important implications on the system's dynamic stability, as will be discussed later. 

\subsection{Enthalpy of formation}

We now discuss the possibility of synthesizing the \ce{Si2X2} and \ce{Si4X2} nanosheets, by analysing the values of the standard enthalpy of formation $\Delta H_{f}^{0}$, obtained by using Eq. \ref{eq_formation_energy} and comparing with the values for pristine silicene structures and graphene, the most well known 2D material. The values are shown in table \ref{table_properties}. 

A negative value of the enthalpy of formation indicates that the formation of a certain compound is exothermic, i.e., the amount of energy it takes to break bonds of the originating species is smaller than the amount of energy released when making the bonds. Herein, a negative $\Delta H_{f}^{0}$ indicates the systems are stable or metastable, since $\Delta H_{f}^{0} < 0$ is a necessary but not sufficient condition for thermodynamic stability \cite{haastrup2018computational}. Nevertheless, it is still possible to synthesize 2D materials by endothermic  processes, i.e., with positive $\Delta H_{f}^{0}$. For instance, the $\Delta H_{f}^{0}$ values of silicene and graphene monolayers are 751 meV/atom  and 70  meV/atom \cite{ipaves2019carbon}, respectively. The $\Delta H_{f}^{0}$ value of silicene obtained here is in good agreement with the available results reported in the literature \cite{revard2016grand,haastrup2018computational}. 

It can be observed in table \ref{table_properties} that the calculated $\Delta H_{f}^{0}$ values for  pure silicene \ce{AA$'$-Si4} and \ce{AB-Si4} bilayers are 609 and 570 meV/atom, respectively, while they are 224 and 220 meV/atom for the \ce{AA$'$A-Si6} and \ce{ABC-Si6} trilayers, respectively. These results confirm the assertion that the FLS synthesis is easier than the silicene monolayer one. Particularly, it has been proposed that it is easier to experimentally produce free-standing 2D materials when the $\Delta H_{f}^{0}$ is under the threshold energy value of 200 meV/atom \cite{ashton2016computational,haastrup2018computational,gjerding2021recent}. 

Among the structures investigated, we found that the silicene bilayer and trilayer sheets doped with substitutional N or P atoms have negative $\Delta H_{f}^{0}$, where the exothermic reaction is likely due to the high reactivity of the silicene surface \cite{ding2013density}. The remaining systems have positive $\Delta H_{f}^{0}$, with values ranging between the threshold (200 meV/atom) and the silicene monolayer enthalpy of formation (751 meV/atom) energies. 

Regarding the \ce{Si2N2} and \ce{Si2P2} bilayers, whose formation are exothermic, the \ce{AA$'$} phase is slightly more favorable than the \ce{AB} one. It is worth mentioning that the $\Delta H_{f}^{0}$ results for \ce{Si2P2} are consistent with other results reported in the literature \cite{huang2015highly}.
The \ce{Si2B2} bilayers are the least energetically favorable systems since they present the largest $\Delta H_{f}^{0}$ values of 648 and 704 meV/atom for \ce{AA$'$} and \ce{AB} stacking, respectively, which is higher than the value found for pristine silicene bilayers. Oppositely, the \ce{Si2Al2} sheets are energetically more favorable than pure silicene bilayers since we found smaller $\Delta H_{f}^{0}$ values of 446 and 404 meV/atom for the \ce{AA$'$} and \ce{AB} stacking configurations.

The doped silicene trilayers present a similar behavior as the bilayers: while the \ce{Si4B2} systems  present the highest positive $\Delta H_{f}^{0}$ among all \ce{Si4X2} structures, the \ce{Si4N2} exhibit the lowest negative $\Delta H_{f}^{0}$ ones. Besides, the silicene trilayers doped with B or Al atoms are energetically less favorable than the pristine silicene trilayers, while the P doped ones present negative $\Delta H_{f}^{0}$ values. 

\subsection{Dynamic stability and elasticity}

To establish the structures' dynamic stability, we used the phonon theory to obtain the respective vibrational spectra. A certain structure is dynamically stable when it has only positive frequencies in the respective phonon dispersion curves. Table \ref{table_properties} presents the stability of all structures investigated here based on that criteria (DS). According to the table, silicene bilayers and trilayers functionalized with P atoms are dynamically stable. On the other hand, silicene structures functionalized with B, Al, or N are dynamically stable only in some stacking configurations.

\begin{figure}[t!]
\centering
{\includegraphics[scale = 0.31, trim={0cm 0.0cm 0cm 0.0cm}, clip]{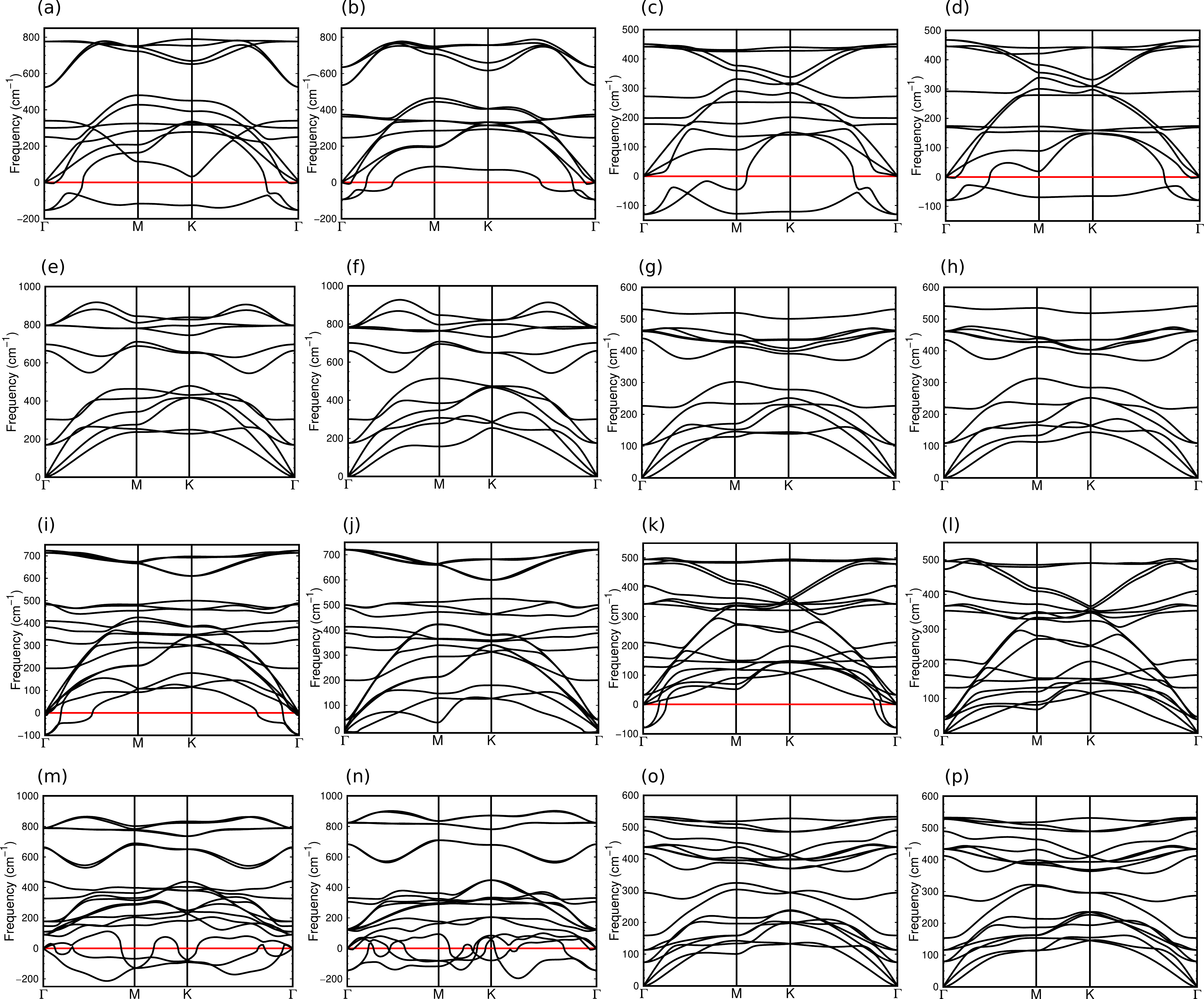}}
\caption{Phonon dispersion curves along the main high symmetry directions of the irreducible Brillouin zone (BZ) of the hexagonal lattice: (a) AA$'$-\ce{Si2B2}, (b) AB-\ce{Si2B2}, (c) AA$'$-\ce{Si2Al2}, (d) AB-\ce{Si2Al2}, (e) AA$'$-\ce{Si2N2}, (f) AB-\ce{Si2N2}, (g)  AA$'$-\ce{Si2P2}, (h) AB-\ce{Si2P2}, (i) \ce{AA$'$A$''$-Si4B2}, (j) \ce{ABC-Si4B2}, (k) \ce{AA$'$A$''$-Si4Al2}, (l) \ce{ABC-Si4Al2}, (m) \ce{AA$'$A$''$-Si4N2}, (n) \ce{ABC-Si4N2}, (o) \ce{AA$'$A$''$-Si4P2}, and (p) \ce{ABC-Si4P2}.}
\label{phonon}
\end{figure}

The structures' phonon dispersion curves are presented in Fig. \ref{phonon}. Both stacking configurations of \ce{Si2B2} and \ce{Si2Al2} bilayers (Fig. \ref{phonon}(a), (b), (c), and (d)) and of \ce{Si4N2} trilayers (Fig. \ref{phonon}(m) and (n)), as well as the \ce{AA$'$A$''$-Si4B2} system (Fig. \ref{phonon}(i)) and \ce{AA$'$A$''$-Si4Al2} trilayer (Fig. \ref{phonon}(k)), present large negative frequencies and, hence, they are dynamically unstable. Moreover, the small negative frequency around the $\Gamma$ valley in the \ce{ABC-Si4B2} system, shown in Fig. \ref{phonon}(j), could indicate that this sheet has lower stability than the other dynamically stable trilayers studied here. This behavior has been found in some theoretical investigations of 2D materials, and has been generally associated with the difficulty in converging the out-of-plane ZA transverse acoustic mode \cite{mohebpour2020prediction,touski2021structural}.

We computed the elastic constants of the systems with Eq. \ref{eq_elastic_energy}, which must satisfy $\rm C_{11}>0$  and $\rm C_{12} < C_{11}$ for elastic stability (Born stability criteria) \cite{mouhat2014necessary}. Moreover, for the dynamically stable (DS) bilayers and trilayers presented in table \ref{table_properties}, we also evaluated the \ce{C44} elastic constant through $\rm C_{44} = \left(C_{11} - C_{12}\right)/2$ expression, the Young modulus $Y^{\rm 2D} = {\rm \left(C_{11}^2 - C_{12}^2\right)/C_{11}}$, and the Poisson ratio $\nu= {\rm  C_{12}/C_{11}}$. The results are displayed in table \ref{table_elastic}, which confirm what was observed in the phonon dispersion spectra shown in figure \ref{phonon}, regarding the stability of the compounds, once all the dynamically stable systems satisfy the Born criteria. For the pristine silicene monolayer, the calculated values for all the elastic constants are in concordance with data from the literature \cite{ding2013density}. Moreover, the \ce{C11} elastic constant of the pure \ce{AB-Si$_4$} bilayer of 121 N/m and \ce{ABC-Si6} trilayer  of 167 N/m  are also in agreement with previously reported results \cite{li2019hydrogenation}. 

\begin{table}[t!]
\begin{center}
\caption{Elastic constants C$_{11}$, C$_{12}$, and C$_{44}$, Young’s modulus $Y^{\rm 2D}$, and Poisson ratio $\nu$ of pristine FLS, Si$_2$X$_2$, and Si$_4$X$_2$ (X = B, N, Al, P) for the dynamically stable configurations. Elastic constants and Young’s modulus are given in N/m and Poisson ratio is dimensionless.}
\begin{tabular}{lllrrrrr}
\hline \hline
 &  &  &  &  & &  &   \\[-3.5mm]
\multicolumn{1}{l}{System}  & \multicolumn{1}{l}{Structure} & \multicolumn{1}{l}{Stacking} &  \multicolumn{1}{c}{C$_{11}$} &  
\multicolumn{1}{c}{C$_{12}$} & \multicolumn{1}{c}{C$_{44}$} &  \multicolumn{1}{c}{ $\,\,Y^{\rm 2D}$} & 
 \multicolumn{1}{c}{$\,\nu$} \\ 
 &  &  &  &  & &  &  \\ [-4mm] 
\hline \hline
 &  &  &  &  & &  &  \\ [-4mm] 
Monolayer~~ & \multicolumn{1}{l}{Si$_2$} & ---&  69 & 19  & 25 & 64 &  0.28   \\  \hline
 &  &  &  &  & &  &  \\ [-4mm] 
\multirow{6}{*}{Bilayer} & \multirow{2}{*}{Si$_4$} & AA$'$ & 124 & 36 & 44  & 113  & 0.29   \\
 &     &  AB   & 121 & 33 & 43 & 111 &  0.28    \\ \cline{2-8}
 &  &  &  &  & &  &  \\ [-4mm] 
 &\multirow{2}{*}{Si$_2$N$_2$}  &  AA$'$  & 296 & 83 & 106 & 273  & 0.28     \\ 
 &      & AB   & 278 & 83 & 97 & 252 & 0.30    \\  \cline{2-8}
 &  &  &  &  & &  &  \\[-4mm]
&\multirow{2}{*}{Si$_2$P$_2$}  &  AA$'$   & 133 & 24 & 55 & 129  & 0.18     \\ 
 &      & AB   & 129 & 23 & 53 & 125  & 0.18     \\ \hline
 &  &  &  &  & &  &  \\[-4mm]
\multirow{6}{*}{Trilayer} & \multirow{2}{*}{Si$_6$} & AA$^{\prime}$A & 157 & 33 & 62 &  150  &  0.21   \\
& & ABC & 167 & 43 & 62 & 156   & 0.25 \\  \cline{2-8}
  &  &  &  &  & &  &  \\[-3.6mm]
 & Si$_4$B$_2$    & ABC  &  198 & 35 & 82  & 192   & 0.17 \\ 
  &  &  &  &  & &  &  \\[-4mm] \cline{2-8}
 &  &  &  &  & &  &  \\[-3.6mm]
  &  Si$_4$Al$_2$ & ABC   & 123  & 84 & 20 & 66 & 0.68     \\ 
  &  &  &  &  & &  &  \\[-4mm] \cline{2-8}
 &  &  &  &  & &  &  \\[-4mm]
& \multirow{2}{*}{Si$_4$P$_2$} & AA$^{\prime}$A$^{{\prime}{\prime}}$ & 182 & 38 & 72 & 174   & 0.21     \\
 & & ABC  & 176 & 36 & 70 & 168   & 0.21 \\ 
\hline  \hline
\label{table_elastic}
\end{tabular}
\end{center}
\end{table}

Additionally, we observe that pristine FLS structures are more stable as they grow in thickness, i.e., the elastic constants and the $Y^{\rm 2D}$ increase with the number of layers \cite{li2019hydrogenation}. For the functionalized FLS, the $Y^{\rm 2D}$ values range from 66 to 273 N/m, which are directly related to the systems' lattice parameter $a$ and intralayer distances $d$. The \ce{Si2N2} bilayers have the smallest $a$ and the largest $Y^{\rm 2D}$, while the \ce{ABC-Si4Al2} trilayer has the largest $a$ and the lowest $Y^{\rm 2D}$. The remaining systems follow the same trend, which has also been observed in other 2D systems \cite{hess2021bonding}. Furthermore, except for the \ce{ABC-Si4Al2}, all the systems investigated have Poisson ratio $\nu$ between 0.17 and 0.30, in the 0-0.5 range  which has been observed in 2D materials \cite{hess2021bonding} and isotropic systems \cite{gercek2007poisson}. Although the $\nu$ value of 0.68 obtained for the \ce{ABC-Si4Al2} trilayer is larger than 0.5, it is similar to values found in stretched silicene monolayer of 0.62 \cite{wang2014stable} and 0.75 \cite{das2018comparison}. Indeed, the Al atom substitution in pristine silicene trilayer increases the lattice parameter and, hence, the Si-Si bond distances.

Our results show that the stability of the binary bilayers depends on the X atoms, as we find that the combination between group-V and group-IV elements, as in  \ce{Si2N2} and \ce{Si2P2} structures, produces dynamically stable systems as they exhibit only positive frequencies and satisfy the Born stability criteria. On the other hand, there is no dynamic stability for the compounds that involve combinations of group-III and group-IV elements, i.e., \ce{Si2B2} and \ce{Si2Al2} bilayers. 

In the trilayers, the stable systems are \ce{ABC-Si4B2}, \ce{ABC-Si4Al2}, \ce{AA$'$A$''$-Si4P2}, and \ce{ABC-Si4P2}, in which the higher structural stability is related to the system with shorter interlayer distances $h_{\rm {Si-Si}}$, which leads to the strong interactions between layers \cite{liu2013structural}. The instability in the \ce{ABC-Si4N2} compound is likely associated with the large Si-Si intralayer bond distance $d_{\rm Si-Si}$ of 2.807 {\AA} and the huge buckling distance $\Delta z_{\rm Si-Si}$ of 2.267 {\AA} in the non-doped intermediate silicene layer. The $d_{\rm Si-Si}$ is larger than the longest Si-Si $\sigma$ bonds (2.697 and 2.7288(15) {\AA}) \cite{kyushin2020silicon} and the huge $\Delta z_{\rm Si-Si}$ is related to the multilayer silicene instability \cite{qian2020multilayer}. As in the \ce{ABC-Si4N2} system, the instability of the \ce{AA$'$A$''$-Si4N2} is also due to the large Si-Si intralayer bond distance $d_{\rm Si-Si}=2.508$ {\AA} and the huge buckling distance $\Delta z_{\rm Si-Si} = 1.861$ {\AA}. Regarding the \ce{AA$'$A$''$-Si4B2} and \ce{AA$'$A$''$-Si4Al2} structures, they present similar behavior since there is a huge negative frequency around the $\Gamma$ valley,  suggesting that a possible synthesis of \ce{Si4B2} and \ce{Si4Al2} trilayers should be in the \ce{ABC} stacking configuration. 
\begin{figure}[bt!]
 \centering
 \vspace{0.2cm}
 \includegraphics[scale = 0.75, trim={0cm 0.0cm 0.0cm 0.0cm}, clip]{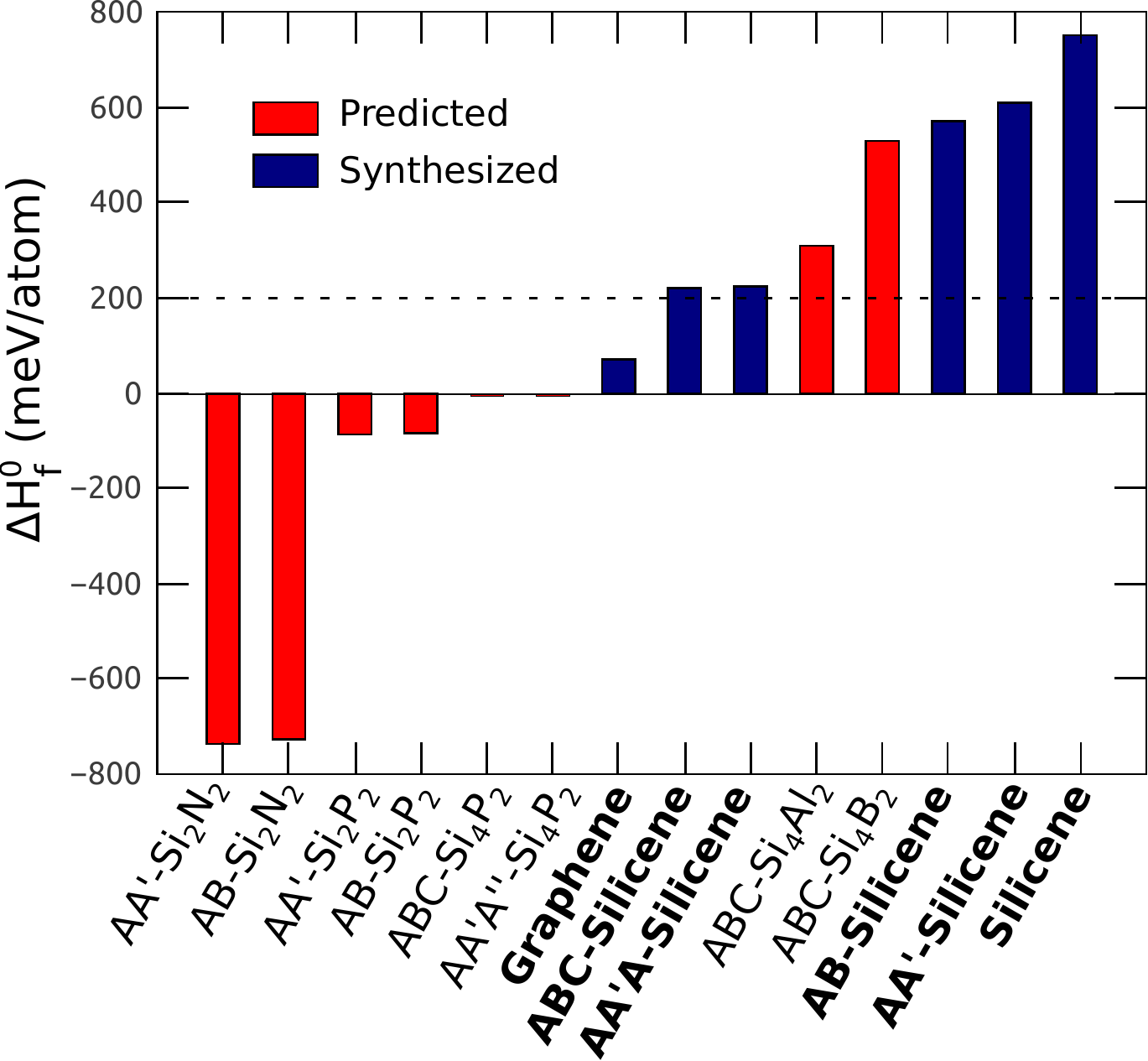} \quad
 \vspace{0.15cm}
\caption{Standard enthalpy of formation, calculated using Eq. \ref{eq_formation_energy}, for the dynamically stable systems. The horizontal dashed line illustrates the empirical threshold energy to experimentally synthesize free-standing 2D materials  \cite{ashton2016computational,haastrup2018computational,gjerding2021recent}.}
\label{formation_energy}
 \end{figure}

Figure \ref{formation_energy} shows the standard enthalpy of formation for all the dynamically stable structures found here. The empirical threshold energy (200 meV/atom) to experimentally synthesize free-standing 2D materials  \cite{ashton2016computational,haastrup2018computational,gjerding2021recent} is also shown. Although FLS has enthalpies of formation greater than 200 meV, they have been free-standing synthesized \cite{liu2018few}. The graphene enthalpy of formation is also included for comparison, which was computed, with respect to graphite, computed using the van der Waals density functional (optB88-vdW) to describe the exchange-correlation energy \cite{ipaves2019carbon}. 
The comparison show the feasibility of experimentally synthesizing, through different techniques, the dynamically stable systems investigated here. 

\subsection{Electronic properties}

\begin{figure}[t!]
  \centering
 \includegraphics[scale = 0.55, trim={0cm 0.0cm 0.0cm 0.0cm}, clip]{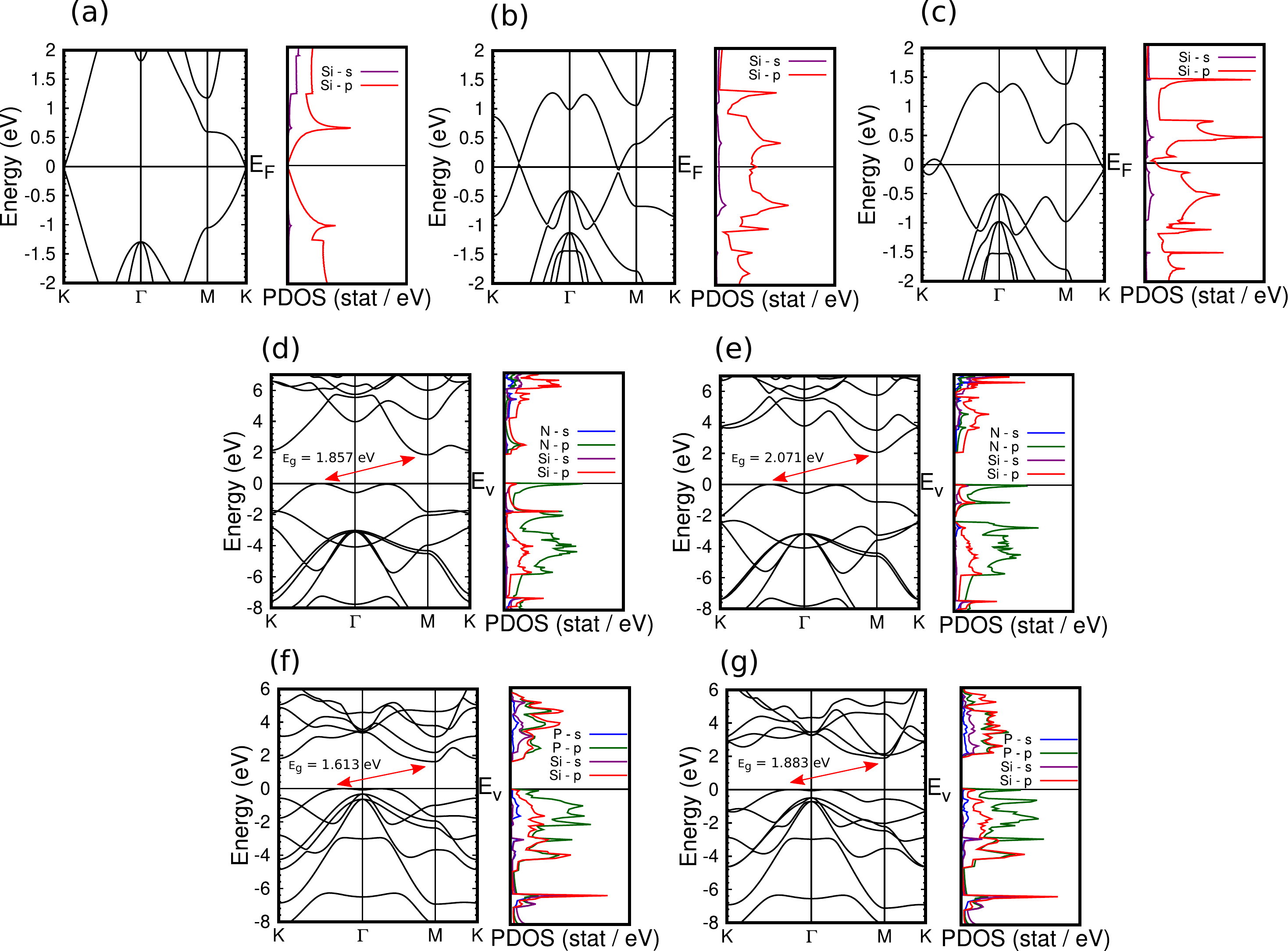}\quad
 \caption{Electronic band structures along the main high-symmetry directions of the BZ (left panels) and projected density of states (PDOS) (right panels) of (a) \ce{Si2}, (b) \ce{AA$'$-Si4}, (c) \ce{AB-Si4}, (d) \ce{AA$'$-Si2N2}, (e) \ce{AB-Si2N2}, (f) \ce{AA$'$-Si2P2}, and (g) \ce{AB-Si2P2}. The PDOS, in units of number of states/eV,  on the Si and X $s$-orbitals are given in purple and blue, respectively, and on the Si and X $p$-orbitals are given in red and green, respectively. E$_{\text{v}}$ represents the valence band top and E$_{\text{F}}$ the Fermi energy.}
 \label{band_bilayer}
 \vspace{0.1cm}
 \end{figure}

\begin{figure}[t!]
  \centering
 \includegraphics[scale = 0.55, trim={0cm 0.0cm 0.0cm 0.0cm}, clip]{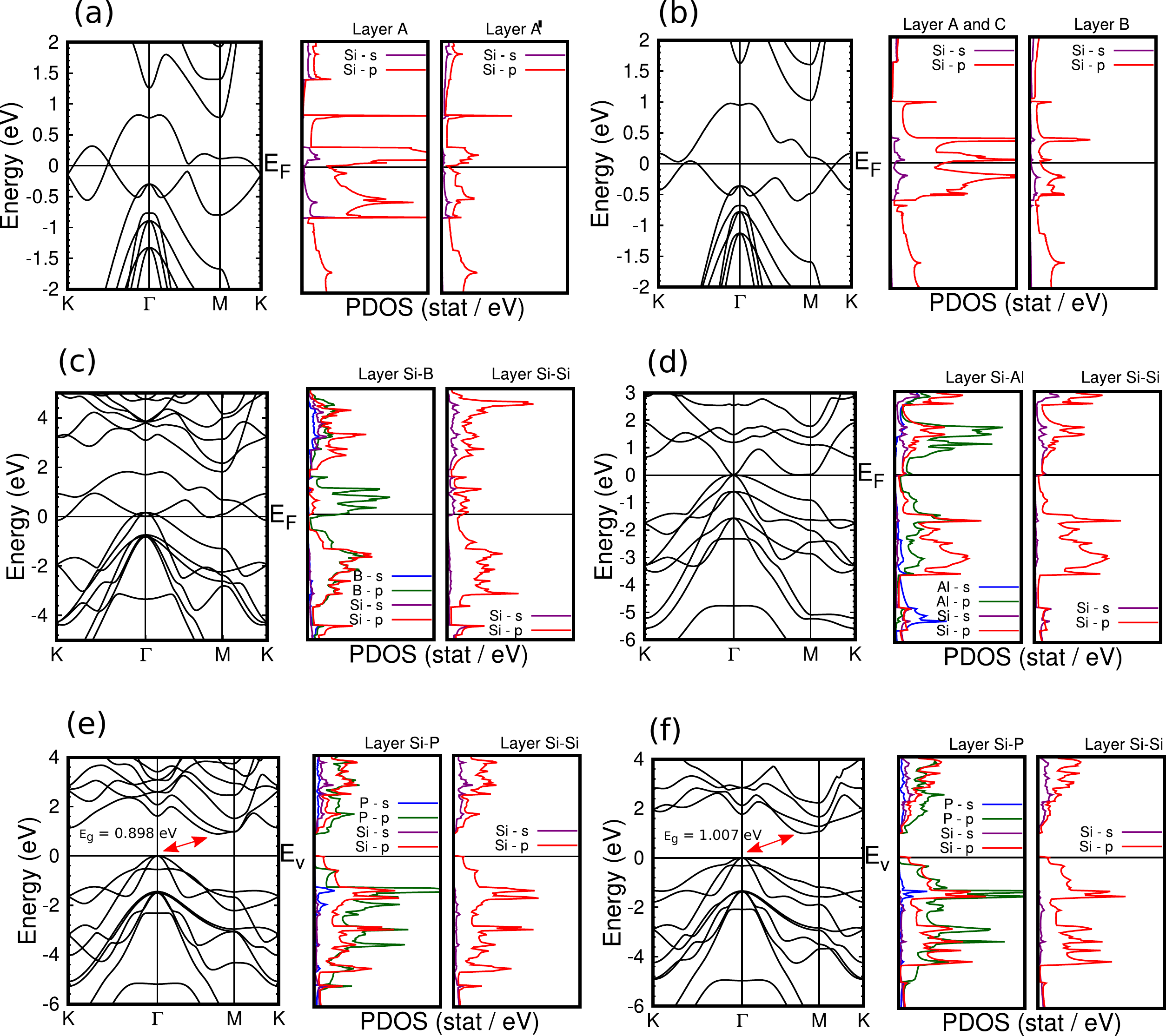}\quad
 \caption{Electronic band structures along the main high-symmetry directions of the BZ (left panels) and projected density of states (PDOS) (right panels) of (a) \ce{AA$'$A-Si6}, (b) \ce{ABC-Si6}, (c) \ce{ABC-Si4B2}, (d) \ce{ABC-Si4Al2}, (e) \ce{AA$'$A$''$-Si4P2}, and (f) \ce{ABC-Si4P2}. The PDOS, in units of number of states/eV, on the Si and X $s$-orbitals are given in purple and blue, respectively, and on the Si and X $p$-orbitals are given in red and green, respectively. E$_{\text{v}}$ represents the valence band top and E$_{\text{F}}$ the Fermi energy.}
 \label{band_trilayer}
 \vspace{0.1cm}
 \end{figure}

Figures \ref{band_bilayer} and \ref{band_trilayer} display the electronic band structures and the projected density of states (PDOS) of the dynamically stable \ce{Si2X2} and \ce{Si4X2} systems (X= B, N, Al, P), as well as the ones of the  pristine silicene monolayer \ce{Si2}, bilayers \ce{Si4}, and trilayers \ce{Si6}. Table \ref{table_electronic} presents the systems' classification and list the values of the indirect electronic band gaps $E_g$, the high-symmetry point or direction of the valence band maximum (VBM) and conduction band minimum (CBM), and their atomic character.  

Previous studies have shown that pure silicene monolayer has zero band gap with a Dirac cone at the K-point in the Brillouin zone \cite{cahangirov2009two} and FLS can present metallic or semiconducting behavior depending on the stacking configuration \cite{padilha2015free,fu2014stacking, qian2020multilayer}. All of our results for pristine silicene structures, displayed in Fig.  \ref{band_bilayer}(a) for \ce{Si2}, Fig. \ref{band_bilayer}(b) and (c) for \ce{Si4}, and Fig. \ref{band_trilayer}(a) and (b) for \ce{Si6}, agree well with the ones reported in the literature, where the investigated FLS are metallic structures. For the functionalized FLS, we found that the \ce{AA$'$-Si2N2}, \ce{AB-Si2N2}, \ce{AA$'$-Si2P2}, and \ce{AB-Si2P2} doped bilayers, Fig. \ref{band_bilayer} (d)-(g), and the \ce{AA$'$A$''$-Si4P2} and \ce{ABC-Si4P2} doped trilayers, Fig.  \ref{band_trilayer}(e) and (f), are indirect band gap semiconductors, while the  \ce{ABC-Si4B2} and \ce{ABC-Si4Al2} trilayers, Fig. \ref{band_trilayer}(c) and (d), are metallic. It should be noticed that these band gap values should be considered as lower limits, as the DFT/vdW is known to underestimate them.

 \begin{table}[hbt]
\begin{center}
\begin{small}
\caption{Electronic properties of the pristine and doped FLS, for the dynamically stable configurations: indirect electronic band gaps $E_g(i)$, given in eV, valence band maximum (VBM) and conduction band minimum (CBM) high-symmetry point or direction (hexagonal BZ), and the atomic character of VBM and CBM. 
These parameters were obtained using the optB88-vdW approximation to describe the exchange-correlation energy. The systems with metallic behavior are also shown.}
\begin{tabular}{llllccc}
\hline \hline
 &  &  &  &  &  & \\[-3.5mm]
\multicolumn{1}{l}{System}  & \multicolumn{1}{l}{Structure} & \multicolumn{1}{l}{Stacking} & \multicolumn{1}{l} {Classification} & \multicolumn{1}{c}{$E_{g}(i)$} & \multicolumn{1}{c}{VBM} (character) &  \multicolumn{1}{c}{CBM} (character)  \\ 
 &  &  &  &  & &  \\ [-4mm] 
\hline \hline
 &  &  &  &  & &  \\ [-4mm] 
Monolayer~~ & \multicolumn{1}{l}{Si$_2$} & --- & semimetal & 0  & ---  & ---     \\  \hline
 &  &  &  &  & &  \\ [-4mm] 
\multirow{6}{*}{Bilayer} & \multirow{2}{*}{Si$_4$} & AA$'$ & metal & --- & --- & ---     \\
 &     &  AB   & metal & --- &  &        \\ \cline{2-7}
 &  &  &  &  & &  \\ [-4mm] 
 &\multirow{2}{*}{Si$_2$N$_2$}  &  AA$'$  & semiconductor  & 1.857 &  K-$\rm\Gamma$ (Si-$p$, N-$p$) & M (Si-$p$, N-$p$)        \\ 
 &      & AB  & semiconductor & 2.071  & K-$\rm\Gamma$ (Si-$p$, N-$p$) & M (Si-$p$)      \\  \cline{2-7}
 &  &  &  &  &  &  \\[-4mm]
&\multirow{2}{*}{Si$_2$P$_2$}  &  AA$'$   & semiconductor  & 1.613 &  K-$\rm\Gamma$ (Si-$p$, P-$p$)  & M (Si-$p$, P-$p$)         \\ 
 &      & AB   &  semiconductor &1.883 &  K-$\rm\Gamma$ (Si-$p$, P-$p$) & M (Si-$p$, P-$p$)          \\ \hline
 &  &  &  &  &  & \\[-4mm]
\multirow{6}{*}{Trilayer} & \multirow{2}{*}{Si$_6$} & AA$^{\prime}$A & metal & --- & --- & ---     \\
& & ABC & metal & --- &  &    \\  \cline{2-7}
  &  &  &  & & &   \\[-3.6mm]
 & Si$_4$B$_2$    & ABC  & metal  & --- & --- & ---   \\ 
  &  &  &  &  & &  \\[-4mm] \cline{2-7}
 &  &  &  &  &  & \\[-3.6mm]
  &  Si$_4$Al$_2$ & ABC   & metal & ---  & --- & ---        \\ 
  &  &  &  &  & &  \\[-4mm] \cline{2-7}
 &  &  &  &  &  & \\[-4mm]
& \multirow{2}{*}{Si$_4$P$_2$} & AA$^{\prime}$A$^{{\prime}{\prime}}$ & semiconductor &0.898 &   $\rm\Gamma$ (Si-$p$) &  $\rm\Gamma$-M (Si-$p$, P-$p$)       \\
 & & ABC  & semiconductor &1.007   & $\rm\Gamma$ (Si-$p$) &  $\rm\Gamma$-M (Si-$p$, P-$p$)    \\ 
\hline  \hline
\label{table_electronic}
\end{tabular}
\end{small}
\end{center}
\end{table}

The differences in the electronic properties result from the distinct \ce{X} doping elements that composed the systems, i.e., the prevailing factor that determined the properties of the compounds is the number of valence electrons of the X atoms. The group-V atoms (N, P) substitution transforms the pristine silicene bilayers and trilayers from metallic to semiconductor. In contrast, the B or Al substitution keeps the metallic behavior of pristine silicene trilayers. Additionally, the Dirac cone in the metallic pristine silicene trilayers, which is below the Fermi energy at the K-point for \ce{AA$'$A} stacking configuration, Fig. \ref{band_trilayer}(a), and in the K-M direction for \ce{ABC} one,  Fig. \ref{band_trilayer}(b) \cite{qian2020multilayer}, is not preserved in the \ce{Si4X2} systems. The electronic band gaps ($E_g$) for \ce{AA$'$-Si2N2}, \ce{AB-Si2N2}, \ce{AA$'$-Si2P2}, \ce{AB-Si2P2}, \ce{AA$'$A$''$-Si4P2}, and \ce{ABC-Si4P2}, are 1.857, 2.071, 1.613, 1.883, 0.898, and 1.007 eV, respectively. The $E_g$ values of \ce{Si2N2} and \ce{Si2P2} functionalized bilayers are in good agreement with reported theoretical results that used the PBE functional approximation for the exchange-correlation energy, values that increased by up to 1 eV (\ce{Si2N2}) and 0.7 eV (\ce{Si2P2}) when the Heyd-Scuseria-Ernzerhof screen-exchange hybrid functional (HSE06) was used \cite{ozdamar2018structural}, suggesting that the electronic band gaps presented in table \ref{table_electronic} are likely underestimated by at least 0.7 eV.

For the \ce{Si2N2} and \ce{Si2P2} semiconducting bilayers, Fig. \ref{band_bilayer} (d)-(g) show that the bottom of the conduction band, at M-point, and the top of the valence band, at the high-symmetry K-$\Gamma$ direction, have contributions mainly from the hybridization of the Si and X $p$-states (X = N, P), with prevailing X $p$-states character. Moreover, the top region of the valence band is highly degenerated with a Mexican-hat dispersion (quartic function), in accordance with other investigations \cite{ozdamar2018structural}. 

The  \ce{AA$'$A$''$-Si4P2} and \ce{ABC-Si4P2} semiconducting trilayers have similar band structures and PDOS, as shown in Fig. \ref{band_trilayer} (e) and (f), with the VBM at the $\Gamma$-point and the CBM at the $\Gamma$-M direction. The PDOS shows that the Si $p$-orbitals  derived from the intermediate (non-doped) Si-Si layer and from the Si-P layer dominate at the VBM, while at the CBM there is also a contribution of the P $p$-states from the Si-P layer. In the \ce{ABC-Si4B2} and \ce{ABC-Si4Al2} metallic systems, Fig. \ref{band_trilayer} (c) and (d), the Fermi level crosses three bands, in the $\Gamma$-point region, that have contributions from the hybridization of the Si and X $p$-orbitals  of the Si-X layer (X = B, Al), and from the Si $s$-state of  the Si-Si layer. Moreover, the Fermi level crossing around the K-point in the \ce{ABC-Si4B2} is mainly derived from the B $p$-states of the Si-B layer and from Si $p$-states of the Si-Si layer. 

These results provide chemical routes to tune and control the FLS electronic structure by increasing the number of layers and, at the same time, doping them to be used in specific  applications. 

%% file: discussion/discussion.tex
{\section{Discussion and Concluding Remarks}}

In summary, we performed a theoretical investigation on  the structural, energetic, dynamic, elastic, and electronic properties of silicene bilayers and trilayers functionalized with group-III and group-V atoms in several stacking configurations. We identified a number of functionalized FLS that are dynamically stable and experimentally accessible, based on the results of the enthalpies of formation, phonon dispersion spectra, and Born stability criteria. Additionally, among them we found that two structures present metallic behavior while the others present semiconducting properties, accordingly to the respective electronic structures, which are directly associated with the atomic species used in the functionalization.      

The \ce{ABC-Si4B2} and \ce{ABC-Si4Al2} metallic systems  present low-buckled surfaces, being the least favorable of the structures studied since they have the greatest positive  enthalpy of formation. However, the enthalpies of formation of these systems are still lower than that of pristine silicene monolayer, which has already been synthesized. In particular, the  \ce{AA$'$A$''$-Si4P2} and \ce{ABC-Si4P2} semiconductor trilayers present an enthalpy of formation nearly zero, i.e., which is lower than the empirical threshold energy of 200 meV/atom for the synthesis of free-standing 2D materials and, therefore, considering the recent synthesis of several 2D materials, the structures studied here are likely to be produced in a near future.  

Further investigations could explore the applicability of the structures propose here. Previous studies have shown that functionalized nanostructures could serve as building blocks for the self-assembly of complex 3D systems \cite{garcia2009,novoselov20162d,ipaves2019carbon}. Particularly, 2D building blocks could be vertically combined to make 2D/3D systems with tailored properties \cite{novoselov20162d,ipaves2019carbon}. Recently, van der Waals heterostructures based on silicon \ce{SiX} (X = N, P, As, Sb, Bi) have been investigated in several stacking configurations, in which the \ce{SiN} was proposed for water oxidation applications and the \ce{SiP} for photocatalytic water splitting \cite{somaiya2021potential}. Therefore, considering the silicon-based structures investigated here, it is interesting to explore if those systems could be used as building blocks and how their properties would modify with different morphologies. The \ce{Si2N2} and \ce{Si2P2} bilayers have been considered for UV-light applications due to their wide band gap \cite{ozdamar2018structural} and as thermoelectric materials at room temperature \cite{somaiya2020exploration}. Within such context, the \ce{AA$'$A$''$-Si4P2} and \ce{ABC-Si4P2} nanosheets  could be explored as thermoelectric materials. Although we found a narrower band gap for these systems in comparison with \ce{Si2P2} bilayers, it is important once more emphasizing  that the band gap is here underestimated due to the DFT/vdW approximation and, hence, we suggest the  \ce{AA$'$A$''$-Si4P2} and \ce{ABC-Si4P2} semiconducting trilayers could be appropriated for  UV-light application as well. 

Moreover, silicene monolayer and FLS nanosheets are promising candidates for energy storage system applications, such as lithium-ion batteries \cite{zhuang2017silicene,liu2018few}. Peculiarly, the silicene monolayer doped with different elements has shown good performance for alkali metal ion batteries (AMIBs). For instance, B/Al doped silicene has low diﬀusion barriers and  higher capacity for sodium-ion batteries (SIBs) and potassium-ion batteries (KIBs) 
 in comparison to pristine silicene \cite{zhu2018potential}. Therefore, since good electrical conductivity is one of the requirements for advantageous electrodes \cite{zhangchallenges}, the \ce{ABC-Si4B2} and \ce{ABC-Si4Al2} metallic structures are promising candidates for AMIBs applications.

%% file: acknowledgement/acknowledgement.tex
\begin{acknowledgement}

This investigation was partially supported by Brazilian federal government agencies CAPES and CNPq. The authors  acknowledge the National Laboratory for Scientific Computing (LNCC/MCTI, Brazil) for providing HPC resources of the Santos Dumont supercomputer \mbox{(http://sdumont.lncc.br)} and Centro Nacional de Processamento de Alto Desempenho em São Paulo (CENAPAD-SP, Brazil). LVCA and JFJ acknowledge support from Brazilian agency CNPq, under projects numbers \mbox{305753/2017-7} and \mbox{305187/2018-0}.
\end{acknowledgement}